## SOFTWARE METAPAPER

# Convenient Interface to Inverse Ising (ConIII): A Python 3 Package for Solving Ising-Type Maximum Entropy Models

Edward D. Lee[1] and Bryan C. Daniels[2]
[1] Department of Physics, 142 Sciences Dr, Cornell University, Ithaca, NY, US
[2] ASU–SFI Center for Biosocial Complex Systems, Arizona State University, Tempe, AZ, US
Corresponding author: Bryan C. Daniels (bryan.daniels.1@asu.edu)

ConIII (pronounced CON-ee) is an open-source Python project providing a simple interface to solving the pairwise and higher order Ising model and a base for extension to other maximum entropy models. We describe the maximum entropy problem and give an overview of the algorithms that are implemented as part of ConIII (https://github.com/eltrompetero/coniii) including Monte Carlo histogram, pseudolikelihood, minimum probability flow, a regularized mean field method, and a cluster expansion method. Our goal is to make a variety of maximum entropy techniques accessible to those unfamiliar with the techniques and accelerate workflow for users.

**Funding Statement:** EDL was supported by an NSF Graduate Fellowship under grant no. DGE-1650441. This research was supported in part by a congressional research grant provided by the Dirksen Congressional Center.

## (1) Overview
**Introduction**
Many biological and social systems are characterized by collective behavior: the correlated pattern of neural firing [1], protein diversity in the immune system [2], conflict participation in monkeys [3], flocking in birds [4], statistics of letters in words [5], or consensus voting in the US Supreme Court [6, 7]. Statistical physics is a natural approach to probing such systems precisely because they are collective [8]. Recently, the development of numerical, analytic, and computational tools have made it feasible in these large collective systems to solve for the maximum entropy (maxent) model that reproduces system behavior, corresponding to solving an "inverse problem." This approach contrasts with the typical problem in statistical physics where one postulates the microscopic model (the Hamiltonian) and works out the physical behavior of the system. In the inverse problem, we find the parameters that correspond to observed behavior of a known system. In many cases, this is a very difficult problem to solve and does not have an analytical solution, and we must rely on analytic approximation and numerical techniques to estimate the parameters.

The pairwise maxent model, the Ising model, has been of particular interest because of its simplicity and generality. A variety of algorithms have been proposed to solve the inverse Ising problem, but different approaches are disparately available on separate code bases in different coding languages, which makes comparison difficult and pedagogy more complicated. With ConIII, it is possible to solve the inverse Ising problem with a variety of algorithms in just a few lines of code.

ConIII is intended to provide a centralized resource for the inverse Ising problem and easy extension to other maxent problems. Although some of the implemented algorithms are specific to the pairwise Ising model, maxent models with arbitrary combinations of higher order constraints can be solved as well by specifying the particular constraints of the maxent model of interest.

In the first few sections of this paper, we give a brief overview of maxent and describe at a high level the algorithms implemented in this package. For those unfamiliar with maxent, we also provide some useful references like [9] and the appendix of [6]. For those seeking more detail about the implemented algorithms, we provide references specific to each algorithm section. Then, we describe the architecture of the package and how to contribute.

**What is maximum entropy?**
Shannon introduced the concept of information entropy in his seminal paper about communication over a noisy channel [9]. Information entropy is the unique measure of uncertainty that follows from insisting on elementary principles of consistency. According to Shannon, the entropy over the probability distribution $p(s)$ of possible discrete configurations $\mathcal{S}$ of a system is

$$S[p] = -\sum_{s \in \mathcal{S}} p(s) \log p(s). \quad (1)$$

These configurations could be on-off patterns of firing in neurons, the arrangement of letters in a word, or the orientation of spins in a material.

When there is no structure in the distribution, meaning that the probability is uniform, entropy is at a maximum. In the context of communication theory as Shannon first discussed, this means that there is no structure to exploit to make a prediction about the next part of an incoming message; thus, maximum entropy means that each new part of the message is maximally "surprising." At the other extreme, when the message consists of the same bit over and over again, we can always guess at the following part of the message and the signal has zero entropy. In the context of modeling, we use entropy not to refer to the difficulty of the message, but to our state of knowledge about it. Entropy precisely measures our uncertainty about the configuration in which we expect to find the system.

Maximum entropy, or maxent, is the formal framework for building models that are consistent with statistics from the data but otherwise as structureless as possible [10, 11]. We begin by choosing a set of $K$ useful or important features from the data $f_k(s)$ that should be true for the model that we are trying to build. These could be whether or not a set of neurons fire together in a temporal bin or the pairwise coincidence for primates in a conflict. The average of this feature across the data set $\mathcal{D}$ with $R$ samples is

$$\langle f_k \rangle_{\text{data}} = \frac{1}{R} \sum_{s \in \mathcal{D}} f_k(s). \quad (2)$$

According to the model in which each observation $s$ occurs with some probability $p(s)$, the same average is calculated over all possible states

$$\langle f_k \rangle = \sum_{s \in \mathcal{S}} p(s) f_k(s). \quad (3)$$

We assert that the model should fit the $K$ features while maximizing entropy. The standard procedure is to solve this by the method of Langrangian multipliers. We construct the Langrangian functional $\mathcal{L}$ by introducing the multipliers $\lambda_k$.

$$\mathcal{L}[p] = -\sum_{s \in \mathcal{S}} p(s) \log p(s) - \sum_{k=1}^{K} \lambda_k \left( \langle f_k \rangle_{\text{data}} - \langle f_k \rangle \right) \quad (4)$$

Then, we solve for the fixed point by taking the derivative with respect to $\lambda_k$. The resulting maxent model is a Boltzmann distribution over states:

$$p(s) = e^{-E(s)} / Z, \quad (5)$$

with relative negative log-likelihood (also known as the energy or Hamiltonian)

$$E(s) = -\sum_{k=1}^{K} \lambda_k f_k(s), \quad (6)$$

and normalization factor (also known as the partition function)

$$Z = \sum_{s \in \mathcal{S}} e^{-E(s)}. \quad (7)$$

Entropy is a convex function of $p$ and the constraints are linear with respect to $p$, so the problem is convex and the maxent distribution unique. Readers familiar with statistical physics will recognize this as an alternative derivation of the microcanonical ensemble, demonstrating that statistical mechanics can be viewed as an inference procedure using the maxent principle [11].

Finding the parameters $\lambda_k$ that match the constraints $\langle f_k \rangle_{\text{data}}$ is equivalent to minimizing the Kullback-Leibler divergence between the model and the data [12]

$$D_{\text{KL}}(p_{\text{data}} || p) = \sum_s p_{\text{data}} \log \left( \frac{p_{\text{data}}(s)}{p(s)} \right) \quad (8)$$

$$\frac{\partial D_{\text{KL}}}{\partial \lambda_k} = \sum_s p_{\text{data}}(s) \frac{\partial (-E(s) - \log Z)}{\partial \lambda_k} = 0 \quad (9)$$
$$\Rightarrow \langle f_k \rangle_{\text{data}} = \langle f_k \rangle.$$

In other words, the parameters of the maxent model are the ones that minimize the information theoretic "distance" to the distribution of the data given the constraints. Note that these parameters are given by the data: once the constraints have been chosen, there is a single maxent solution, with no free parameters.

### The Ising model

The Ising model is a statistical physics model of magnetism, also known as the pairwise maxent model [13]. It consists of a set of spins $\{s_i\}$ with 2 possible orientations (up and down), each responds to its own external magnetic field $h_i$ and each pair is coupled to each other with pairwise coupling $J_{ij}$. The strength of the magnetic field determines the tendency of each of the spins to orient in a particular direction and the couplings determine whether the spins tend to point together ($J_{ij} > 0$) or against each other ($J_{ij} < 0$). Typically, neighbors are defined as spins that interact with one another given by some underlying network structure. **Figure 1** shows a fully-connected example.

The energy of each configuration determines its probability via Eq (5),

$$E(s) = -\sum_{i<j}^{N} J_{ij} s_i s_j - \sum_{i=1}^{N} h_i s_i, \quad (10)$$

such that lower energy states are more probable.

We can derive the Ising model from the perspective of maxent. Fixing the the means and pairwise correlations to those observed in the data

$$\langle s_i \rangle = \langle s_i \rangle_{\text{data}} \quad (11)$$

$$\langle s_i s_j \rangle = \langle s_i s_j \rangle_{\text{data}} \quad (12)$$

we go through the procedure of constructing the Langrangian from Eq 4

$$\mathcal{L}[p] = -\sum_s p(s) \log p(s) + \sum_{i<j}^{N} J_{ij} \langle s_i s_j \rangle + \sum_i^{N} h_i \langle s_i \rangle \quad (13)$$

$$\frac{\partial \mathcal{L}[p]}{\partial p(s)} = -\log p(s) - 1 + \sum_{i<j}^{N} J_{ij} s_i s_j + \sum_i^{N} h_i s_i \quad (14)$$

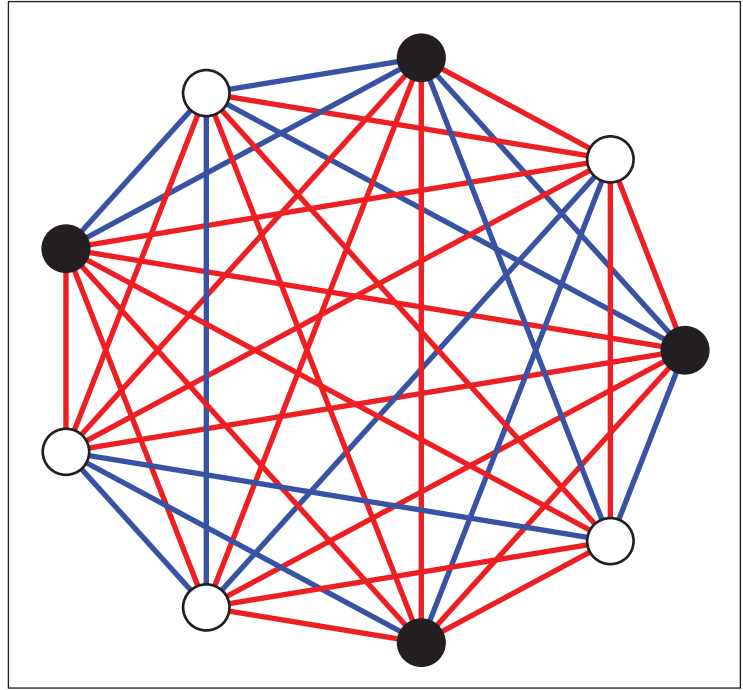

**Figure 1:** Example of a fully connected pairwise Ising model with positive and negative couplings. Each spin $s_i$ (circle) can take one of two states (black or white, corresponding to −1 and 1) and is connected to every other spin in the system with a positive (red) or negative (blue) coupling. These states could describe the on-off patterns of firing in neurons, the orientation of spins in a material, or if each spin is no longer binary the arrangement of letters in a word (a Potts model).

$$\log p(s) = -1 + \sum_{i<j}^{N} J_{ij} s_i s_j + \sum_i^N h_i s_i \qquad (15)$$

$$p(s) = e^{-E(s)}/Z \qquad (16)$$

where the −1 in Eq 15 has been absorbed into the normalization factor

$$Z = \sum_s e^{-E(s)}. \qquad (17)$$

such that the probability distribution is normalized $\sum_s p(s) = 1$. Thus, the resulting model is exactly the Ising model mentioned earlier.

Despite the simplicity of the Ising model, the structure imposed by the discrete nature of the spins means that finding the parameters is challenging analytically and computationally. In the last few years, numerous techniques have been suggested for solving the inverse Ising problem exactly or approximately [14]. We have implemented some of them in ConIII and designed a package structure to make it easily extensible to include more methods. Here, we briefly describe the algorithms that are part of the first official version of the package. The goal is to give the user a sense of how they work without getting bogged down in heavy detail. For more detail, we suggest perusing the papers referenced in each section or the review [14]. For a complete beginner, it may be useful to first get familiar with a slower introduction like in the Appendices of Ref [6], Ref [15], or Ref [10].

**Inverse Ising methods implemented in ConIII**
*Enumeration*
The naïve approach that only works for small systems is to write out the equations from Eq 9 and solve them numerically. After writing out all $K$ equations,

$$\langle f_k \rangle = -\frac{\partial \ln Z}{\partial \lambda_k} = \langle f_k \rangle_{\text{data}}, \qquad (18)$$

we can use any standard root-finding algorithm to find the parameters $\lambda_k$. This approach, however, involves enumerating all states of the system, whose number grows exponentially with system size.

For the Ising model, writing down the equations has a number of steps $\mathcal{O}(K^2 2^N)$, where $K$ is the number of constraints and $N$ the number of spins. Each evaluation of the objective in the root-finding algorithm will be of the same order. For relatively small systems, around $N \leq 15$, this approach is feasible on a typical desktop computer and is a good way to test the results of a more complicated algorithm. This approach is implemented by the `Enumerate` class.

*Monte Carlo Histogram (MCH)*
Perhaps the most straightforward and most expensive computational approach is Monte Carlo Markov Chain (MCMC) sampling. A series of states sampled from a proposed $p(s)$ is produced by MCMC to approximate $\langle f_k \rangle$ and determine how close we are to matching $\langle f_k \rangle_{\text{data}}$. The parameters are then adjusted using a learning rule, and both sampling and learning are repeated until a stopping criterion is met. This can be combined with a variety of approximate gradient descent methods to reduce the number of sampling steps by predicting how the distribution will change if we modify the parameters slightly. The particular technique implemented in ConIII is the Monte Carlo Histogram (MCH) method [16].

Since the sampling step is expensive, the idea behind MCH is to reuse a sample for more than one gradient descent step [16]. Given that we have a sample with probability distribution $p(s)$ generated with parameters $\lambda_k$, we would like to estimate the proposed distribution $p'(s)$ from adjusting our parameters $\lambda'_k = \lambda_k + \Delta\lambda_k$. We can leverage our current sample to make this extrapolation.

$$p' = \frac{p'}{p} p \qquad (19)$$

$$p'(s) = \frac{Z}{Z'} e^{\sum_k \Delta\lambda_k f_k(s)} p(s) \qquad (20)$$

To estimate the average,

$$\sum_s p'(s) f_k(s) = \frac{Z}{Z'} \sum_s p(s) e^{\sum_k \Delta\lambda_k f_k(s)} f_k(s) \qquad (21)$$

To be explicit about the fact that we only have a sampled approximation to $p$, we replace $p$ with the sample distribution.

$$\langle f_k \rangle' = \frac{Z}{Z'} \left\langle e^{\sum_k \Delta\lambda_k f_k(s)} f_k(s) \right\rangle_{sample} \qquad (22)$$

Likewise, the ratio of the partition function can be estimated

$$\frac{Z}{Z'} \approx 1 \bigg/ \left\langle e^{\sum_k \Delta\lambda_k f_k(s)} \right\rangle_{sample} \qquad (23)$$

At each step, we update the Lagrangian multipliers $\{\lambda_k\}$ while being careful to stay within the bounds of a reasonable extrapolation. One suggestion is to update the parameters with some inertia [17]

$$\Delta\lambda_k(t+1) = \Delta\lambda_k(t) + \epsilon \Delta\lambda_k(t-1) \qquad (24)$$

$$\Delta\lambda_k(t) = \eta \left( \langle f_k \rangle' - \langle f_k \rangle \right) \qquad (25)$$

This has a fixed point at the correct parameters.

In practice, MCH can be difficult to tune properly and one must check in on the progress of the algorithm often. One issue is choosing how to set the learning rule parameters $\eta$ and $\epsilon$. One suggestion for $\eta$ is to shrink it as the inverse of the number of iterations [17]. Another issue is that parameters cannot be changed by too much when using the MCH approximation step or the extrapolation to $\lambda'_k$ will be inaccurate and the algorithm will fail to converge. In ConIII, this can be controlled by setting a bound on the maximum possible change in each parameter $\Delta\lambda_{max}$ and restricting the norm of the vector of change in parameters $\sum_k \sqrt{\Delta\lambda_k^2}$. Another issue is setting the parameters of the MCMC sampling routine. Both the burn time (the number of iterations before starting to sample) and sampling iterations (number of iterations between samples) must be large enough that we are sampling from the equilibrium distribution. Typically, these are found by measuring how long the energy or individual parameter values remain correlated as MCMC progresses. The parameters may need to be updated during the course of MCH because the sampling parameters may need to change with the estimated parameters of the model. For some regimes of parameter space, samples are correlated over long times and alternative sampling methods like Wolff or Swendsen-Wang would vastly reduce time to reach the equilibrium distribution although these are not included in the current release of ConIII. We do not discuss these sampling details here, but see Refs [18, 19] for examples.

The main computational cost for MCH lies in the sampling step. For each iteration of MCH, the runtime is proportional to the number of samples $n$, number of MCMC iterations $T$, and the number of constraints for the Ising model $N^2$, $\mathcal{O}(TnN^2)$, whereas the MCH estimate is relatively quick $\mathcal{O}(tnN^2)$ because the number of MCH approximation steps needed to converge is much smaller than the number of MCMC sampling iterations $t \ll T$.

MCH is implemented in the `MCH` class.

### Pseudolikelihood

The pseudolikelihood approach is an analytic approximation to the likelihood that drastically reduces the computational complexity of the problem and is exact as $N \to \infty$ [20]. We calculate the conditional probability of each spin $s_i$ given the rest of the system $\{s_{j \neq i}\}$

$$p(s_i | \{s_{j \neq i}\}) = \left( 1 + e^{-2s_i \left( h_i + \sum_{j \neq i} J_{ij} s_j \right)} \right)^{-1} \qquad (26)$$

Taking the logarithm, we define the approximate log-likelihood by summing over data points indexed by r:

$$f(h_i, \{J_{ij}\}) = \sum_{r=1}^R \ln p\left(s_i^{(r)} | \{s_{j \neq i}\}^{(r)}\right). \qquad (27)$$

In the limit where the ensemble is well sampled, the average over the data can be replaced by an average over the ensemble:

$$f(h_i, \{J_{ij}\}) = \sum_s \ln p(s_i | \{s_{j \neq i}\}) p(s; h_i, \{J_{ij}\}). \qquad (28)$$

To find the point of maximum likelihood for a single spin $s_i$, we calculate the analytical gradient and Hessian, $\partial f / \partial J_{ij}$ and $\partial^2 f / \partial J_{ij} \partial J_{i'j'}$, for a Newton conjugate-gradient descent method. After maximizing likelihood for all spins, the maximum likelihood parameters may not satisfy the symmetry $J_{ij} = J_{ji}$. We impose the symmetry by insisting that

$$J'_{ij} = (J_{ij} + J_{ji}) / 2. \qquad (29)$$

Pseudolikelihood is extremely fast and often surprisingly accurate. Each calculation of the gradient is order $\mathcal{O}(RN^2)$ and Hessian $\mathcal{O}(RN^3)$, which must be done for all $N$. With analytic forms for the gradient and Hessian, the conjugate-gradient descent method tends to converge quickly.

Pseudolikelihood for the Ising model is implemented in `Pseudo`.

### Minimum Probability Flow (MPF)

Minimum probability flow involves analytically approximating how the probability distribution *changes* as we modify the *configurations* [21, 22]. In the methods so far mentioned, the approach has been to maximize the objective (the likelihood function) by immediately taking the derivative with respect to the parameters. With MPF, we first posit a set of dynamics that will lead the data distribution to equilibrate to that of the model. When these distributions are equivalent, then there is no "probability flow" between them. This technique is closely related to score matching, where we instead have a continuous state space and can directly take the derivative with respect to the states without specifying dynamics [23].

First note that Monte Carlo dynamics (satisfying ergodicity and detailed balance) would lead to equilibration to the stationary distribution. The dynamics are specified by a transition matrix, an example of which is given in Ref [22]:

$$\dot{p}_s = \sum_{s'\neq s} \Gamma_{ss'} p_{s'} - \sum_{s'\neq s} \Gamma_{s's} p_s \qquad (30)$$

$$\Gamma_{ss'} = g_{ss'} \exp\left[\frac{1}{2}(E_{s'} - E_s)\right] \qquad (31)$$

with transition probabilities $\Gamma_{ss'}$ from state s′ to state s. The connectivity matrix $g_{ss'}$ specifies whether there is edge between states s and s′ such that probability can flow between them. By choosing a sparse $g_{ss'}$ while not breaking ergodicity, we can drastically reduce the computational cost of computing this matrix.

Imagine that we start with the distribution over the states as given by the data and run the Monte Carlo dynamics. When data and model distributions are different, probability will flow between them and indicate that the parameters must be changed. By minimizing a derivative of the Kullback-Leibler divergence, we measure how the difference between the model and the states in the data $\mathcal{D}$ changes when the dynamics are run for an infinitesimal amount of time.

$$L(\{\lambda_k\}) \equiv \partial_t D_{KL}(p^{(0)} || p^{(t)}(\{\lambda_k\})) = \sum_{s\notin\mathcal{D}} \dot{p}_s(\lambda_k) \qquad (32)$$

The idea is that this derivative is also minimized with optimal parameters: the MPF algorithm looks for a minimum of the objective function $L$.

For the Ising model, each evaluation of the objective function where $\Gamma_{ss'}$ connects each data state with $G$ neighbors has runtime $\mathcal{O}(RGN^2)$. In a large fully connected system, $G \sim 2^N$ would be prohibitively large so a sparse choice is necessary.

MPF is implemented in the `MPF` class.

### Regularized mean-field method

One attractively simple and efficient approach uses a regularized version of mean-field theory. In the inverse Ising problem, mean-field theory is equivalent to treating each binary individual as instead having a continuously varying state (corresponding to its mean value). The inverse problem then turns into simply inverting the correlation matrix $C$ [24]:

$$J_{ij}^{\text{mean-field}} = -\frac{(C^{-1})_{ij}}{\sqrt{p_i(1-p_i)p_j(1-p_j)}}, \qquad (33)$$

where

$$C_{ij} = \frac{p_{ij} - p_i p_j}{\sqrt{p_i(1-p_i)p_j(1-p_j)}}, \qquad (34)$$

and where $p_i$ corresponds to the frequency of individual i being in the active (+1) state and $p_{ij}$ is the frequency of the pair i and j being simultaneously in the active state.

A simple regularization scheme in this case is to discourage large values in the interaction matrix $J_{ij}$. This corresponds to putting more weight on solutions that are closer to the case with no interactions (independent individuals). A particularly convenient form adds the following term, quadratic in $J_{ij}$, to the negative log-likelihood:

$$\gamma \sum_i \sum_{i<j} J_{ij}^2 p_i(1-p_i) p_j(1-p_j). \qquad (35)$$

In this case, the regularized version of the mean-field solution in (33) can be solved analytically, with the slowest computational step coming from the inversion of the correlation matrix. For details, see Refs. [3, 25].

The idea is then to vary the regularization strength γ to move between the non-interacting case (γ → ∞) and the naively calculated mean-field solution (33) (γ → 0). While there is no guarantee that varying this one parameter will produce solutions that are good enough to "fit within error bars," this approach has been successful in at least one case of fitting social interactions [3].

The inversion of the correlation matrix is relatively fast, bounded by $\mathcal{O}(N^3)$. Finding the optimal γ involves Monte Carlo sampling from the model distribution, which has computational cost similar to MCH. It is, however, much more efficient because we are only optimizing a single parameter.

This is implemented in `RegularizedMeanField`.

### Cluster expansion

Adaptive cluster expansion [24, 25] iteratively calculates terms in the cluster expansion of the entropy $S$:

$$S - S_0 = \sum_\Gamma \Delta S_\Gamma, \qquad (36)$$

where the sum is over clusters Γ and in the exact case includes all $2^N - 1$ possible nonempty subsets of individuals in the system. In the simplest version of the expansion, one expands around $S_0 = 0$. In some cases it can be more advantageous to expand around the independent individual solution or one of the mean-field solutions described in the previous section [25].

The inverse Ising problem is solved independently on each of the clusters, which can be done exactly when the clusters are small. These results are used to construct a full interaction matrix $J_{ij}$. The expansion starts with small clusters and expands to use larger clusters, neglecting any clusters whose contribution $\Delta S_\Gamma$ to the entropy falls below a threshold. To find the best solution that does not overfit, the threshold is initially set at a large value and then lowered, gradually including more clusters in the expansion, until samples from the resulting $J_{ij}$ fit the desired statistics of the data sufficiently well.

The runtime will depend on the size of clusters included in the expansion. If the expansion is truncated at clusters of size $n$, the worst-case runtime would be $o\left(\binom{N}{n} 2^n\right)$. The point is that $S$ can often be accurately estimated even when $n \ll N$. The adaptive cluster expansion method is implemented in the `ClusterExpansion` class.

### Implementation and architecture

The package is divided into three principal modules containing the algorithms for solving the inverse maxent problem (`solvers.py`), the Monte Carlo Markov Chain (MCMC) sampling algorithms (`samplers.py`), and supporting "utility" functions for the remaining modules (`utils.py`) as shown in **Figure 2**. Besides the `utils.py` module, the package is organized around classes that correspond to different algorithms. This class-based structure ensures that the state of the solver or sampler, including the data it was fit to and the current guess for the parameters, are all contained within the instance of the algorithm class. As a result, the current state of work can be saved and moved between workstations using the Python package `dill`.

For the solvers, the different algorithms available are accessible from the `coniii.solvers` module as listed in **Figure 2**. These algorithm classes are all derived from a base `Solver` class as shown in **Figure 2**. The module `Solver.solve` serves as the interface for solving the inverse maxent problem. To keep the solution algorithms generic enough to solve a variety of different maxent problems, they all require that the user define the maxent model upon instantiation through the definition of keyword arguments like `calc_observables`. The particular methods required to specify the maxent problem differ by algorithm, but for the pairwise maxent problem we have made it easy by defining those functions as part of the package. These helper functions are available as part of the `utils.py` module and their use is demonstrated in the Jupyter notebook usage guide.

The MCMC sampling algorithms are likewise based on a class architecture derived from `Sampler` as shown in **Figure 2**. Each instance of `Solver` automatically instantiates this class under `Solver.sampler` and wraps calls to it. For the Ising model, this is an instance of `Metropolis`. Other sampling algorithms listed in the `samplers.py` box in **Figure 2** will be released with later versions of this package.

**Quality Control**

For checks of basic functionality, the package is released with unit tests that can be run with the Python package pytest.

The most direct test of the algorithms is to generate a system where the parameters are known, sample from the system to generate a data set, and run the inverse solution to make sure that the correlations and parameters match the known values. With a finite sample, exact correspondence to the correct parameters is not expected although differences should decrease with a larger sample. Furthermore, most of the algorithms only return an approximate solution such that the fidelity of the found parameters to the original ones will depend on the sample size and whether or not the approximation is valid. The Jupyter notebook released with the software provides examples for using the algorithms included in ConIII for a random system of five spins. We recommend that the user run this notebook to check how well different algorithms converge to the solutions depending on the algorithm and sample size.

More importantly, the user can check if the algorithms match the expected correlations closely or not. How one checks the validity of a particular maxent model for data is beyond the scope of this paper, but we point the reader to the appendix of Ref [6] where the methodology is explained in detail for a broad audience.

If there are any issues or bugs in the software, we organize improvements and patches through the GitHub repository where both issues can be filed and pull requests made.

## (2) Availability

**Operating system**
Linux, MacOS, Windows

**Programming language**
Python 3.6, 3.7

**Dependencies**
Python packages multiprocess ≥ v0.70.5 and <v1, numpy, scipy, joblib, matplotlib, numba ≥ v0.39.0, dill.

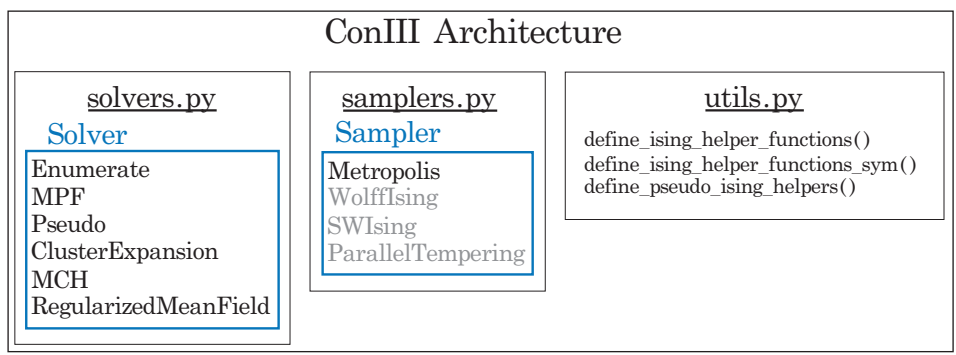

**Figure 2:** Brief summary of ConIII architecture. The principal modules are `solvers.py`, `samplers.py`, and `utils.py`. The module `solvers.py` contains classes based on Solver that each implement a different algorithm for solving the relevant inverse maxent problem accessible through the method `Solver.solve()`. The `samplers.py` module contains the Metropolis algorithm for Monte Carlo Markov Chain sampling and will support other samplers in future versions (gray font) including Wolff sampling, Swendsen-Wang sampling, and parallel tempering. The `utils.py` module contains supporting functions for the other modules such as the few examples listed. ConIII's modularized structure ensures that contributed algorithms can be appended independently of existing code.

**Software location**
*Name:* PyPI
*Persistent identifier:* https://pypi.org/project/coniii/
*Licence:* MIT License
*Publisher:* Edward D. Lee
*Version published:* v1.1.4
*Date published:* 1/6/2019

*Code repository*
*Name:* GitHub
*Persistent identifier:* https://github.com/eltrompetero/coniii
*Licence:* MIT License
*Version published:* v1.1.4

*Name:* Zenodo
*Persistent identifier:* https://doi.org/10.5281/zenodo.2236632
*Licence:* MIT License
*Version published:* v1.1.1

**Language**
English

## (3) Reuse potential

To contribute either an algorithm for the inverse maxent problem or a sampling technique, we suggest following the template for the classes described in the base `Solver` and `Sampler` classes. New algorithms should be filed as a pull request to the GitHub repository along with an example solution that can be included in the usage guide Jupyter notebook and unit tests.

Documentation for the package is included as part of the GitHub repository and also hosted online at https://eddielee.co/coniii/index.html.

**Acknowledgements**
We thank the anonymous reviewers for helpful feedback on both the manuscript and the accessibility of the software.

**Competing Interests**
The authors have no competing interests to declare.